\documentclass[conference]{IEEEtran}

\newif\ifanonymous
\anonymousfalse % Set to \anonymousfalse for non-anonymous mode

\IEEEoverridecommandlockouts
\usepackage{packages}
\def\BibTeX{{\rm B\kern-.05em{\sc i\kern-.025em b}\kern-.08em
    T\kern-.1667em\lower.7ex\hbox{E}\kern-.125emX}}
\begin{document}

%to decide
%\title{Tracing Debt: Historical Analysis of Self-Admitted Technical Debt in ML-Enabled Systems}
\title{A First Look at the Lifecycle of DL-Specific Self-Admitted Technical Debt}

%\title{Duration Matters: A Preliminary Study of DL-Specific Self-Admitted Technical Debt in ML-Enabled Systems}
%\title{Preliminary Analysis on the Lifecycle of Self-Admitted Technical Debt in ML-enabled Systems}
\ifanonymous
    \author{Anonymous Authors} % Leave blank for anonymous mode
\else
    \author{
    \IEEEauthorblockN{Gilberto Recupito, Vincenzo De Martino, Dario Di Nucci, Fabio Palomba}
    \IEEEauthorblockA{SeSa Lab - University of Salerno, Fisciano, Italy\\
    \{grecupito, vdemartino, ddinucci, fpalomba\}@unisa.it}
}
\fi

\maketitle

\begin{abstract}
%Self-Admitted Technical Debt (SATD) in Machine Learning (ML) and Deep Learning (DL)-enabled systems is an increasing concern, significantly impacting software maintainability, performance, and quality. Despite its critical impact, the lifecycle of DL-specific SATD in these systems remains underexplored, particularly regarding how developers introduce, acknowledge, and address SATD over time. 
The rapid adoption of Deep Learning (DL)-enabled systems has revolutionized software development, driving innovation across various domains. However, these systems also introduce unique challenges, particularly in maintaining software quality and performance. Among these challenges, Self-Admitted Technical Debt (SATD) has emerged as a growing concern, significantly impacting the maintainability and overall quality of ML and DL-enabled systems. Despite its critical implications, the lifecycle of DL-specific SATD—how developers introduce, acknowledge, and address it over time—remains underexplored.
This study presents a preliminary analysis of the persistence and lifecycle of DL-specific SATD in DL-enabled systems. The purpose of this project is to uncover the patterns of SATD introduction, recognition, and durability during the development life cycle, providing information on how to manage these issues. Using mining software repository techniques, we examined 40 ML projects, focusing on 185 DL-specific SATD instances. The analysis tracked the introduction and persistence of SATD instances through project commit histories to assess their lifecycle and developer actions. The findings indicate that DL-specific SATD is predominantly introduced during the early and middle stages of project development. Training and Hardware phases showed the longest SATD durations, highlighting critical areas where debt accumulates and persists. Additionally, developers introduce DL-specific SATD more frequently during feature implementation and bug fixes. This study emphasizes the need for targeted DL-specific SATD management strategies in DL-enabled systems to mitigate its impact. By understanding the temporal characteristics and evolution of DL-specific SATD, developers can prioritize interventions at critical stages to improve the maintainability and quality of the system.
%This study presents a preliminary analysis of the evolution of SATD in ML-enabled systems, focusing on how developers introduce, recognize, and address SATD over time. Specifically, we aim to provide insights into the developers' introduction and recognition of SATD and its persistence throughout its lifecycle. By leveraging Mining Software Repositories, our analysis examines the evolution of 40 ML projects and 185 SATD instances. The findings reveal a notable persistence of SATD, primarily introduced during the early stages of project evolution. This study highlights the critical need to manage the impact of SATD on ML-enabled system evolution and offers initial insights into the temporal characteristics of SATD instances.
\end{abstract}


\section{Introduction}\label{Intro}

Deep Learning (DL) has rapidly established itself as a transformative force across diverse industries. By leveraging advanced neural network architectures, DL provides powerful solutions to complex challenges, revolutionizing fields such as healthcare, autonomous driving, and natural language processing \cite{ahmad2023revolutionizing,modelsaiCompaniesUsing,chib2023recent}. DL is a subfield of Machine Learning (ML), a broader field focused on developing algorithms that learn from data \cite{lecun2015deep,goodfellow2016deep}. ML techniques, including DL, are increasingly integrated into traditional software systems, resulting in ML-enabled systems that incorporate at least one ML component \cite{martinez2022software}.
%For example, AI chatbots like ChatGPT\footnote{\url{https://openai.com/index/chatgpt/}} utilize large language models (LLMs) to engage users in dynamic and meaningful conversations.
Although the adoption of these technologies in software systems has spread significantly to provide novel services, they present unique challenges that can increase the effort required to maintain and ensure the high quality of the architecture of ML-enabled systems \cite{nazir2024architecting}.
One of the main challenges that can negatively impact the quality of software systems, particularly DL-enabled systems, is related to technical debt \cite{bhatia2023empirical}. This term refers to the issues arising when developers opt for quick fixes or suboptimal solutions during the development of machine learning models \cite{sculley2015hidden,mcguire2023sustainability}.
Over time, accumulating technical debt can lead to significant quality degradation, affecting not only maintainability but also other aspects of DL-specific quality such as performance and system security\cite{recupito2024technical}.

Developers acknowledge the presence of these quality issues explicitly, often through Self-Admitted Technical Debt (SATD) \cite{potdar2014exploratory}, which highlights potential problems in source code comments or issue trackers \cite{li2023automatic}. Various methods have been proposed to help developers manage SATD, noting that comments usually include recognizable keywords such as \emph{`TODO'} and \emph{`FIXME'} \cite{mastropaolo2023towards}. 
An example of introducing an SATD is shown in the Listing \ref{list:example}. Developers add comments to raise an issue and to describe the suboptimal solution they employed, using the keyword (\ie \quoted{\texttt{TODO}}) to warn of the need for further rework in the future, establishing a debt.

%In the context of DL, raised issues by SATD can affect all stages of the development pipeline, including data preprocessing, model training, and system deployment \cite{obrien202223}.

%The code snippet contains SATD, explicitly acknowledged in the comment: \#TODO: use tensor inputs instead of PIL. The developer prioritizes maintaining compatibility with existing functionality over switching to tensor-based inputs, postponing the improvement and creating a trade-off that reflects technical debt.

% Define custom colors
\definecolor{bggray}{rgb}{0.95, 0.95, 0.95}
\definecolor{highlight}{rgb}{0.85, 1.0, 0.85}
\definecolor{darkgreen}{rgb}{0.0, 0.5, 0.0}

\begin{minipage}[c]{0.96\columnwidth}
\label{list:example}
\begin{lstlisting}[caption=Example of the introduction of a SATD.]
def get_dummy_inputs(self, device, seed=0):
# TODO: use tensor inputs instead of PIL, this is here just to leave the old expected_slices untouched
    image = floats_tensor((1, 3, 32, 32), rng=random.Random(seed)).to(device)@@
    image=image.cpu().permute(0, 2, 3, 1)[0]
    init_image = Image.fromarray(np.uint8(image)).convert("RGB").resize((64, 64))
    mask_image = Image.fromarray(np.uint8(image + 4)).convert("RGB").resize((64, 64))
\end{lstlisting}
\end{minipage}

%While previous work has defined a taxonomy of DL-specific SATD, to the best of our knowledge, there is no specific work aimed at automatically resolving 
%Continuare descrivendo (1) i Dl-specific SATD (2) il gap attuale (3) il lavoro svolto basato sul lavoro di Pepe (4) i risultati e (5) le conclusioni

Therefore, DL-specific SATD presents unique challenges compared to conventional SATD because of the distinct characteristics of DL components. Developers must consider SATD from conventional systems and those specific to ML and DL software, which are particularly vulnerable to the accumulation of technical debt \cite{sculley2015hidden}. One notable example is the work conducted by Pepe et al. \cite{pepe2024taxonomy}, which defined a taxonomy of 41 distinct types of DL-specific SATD specific to DL-enabled systems. These types of SATD encompass various phases of DL, including data, model, training, and inference.

While previous efforts have defined DL-specific SATD in ML projects by creating a taxonomy, they have not thoroughly examined the life cycle of these technical debts. Examining when SATD instances are introduced and determining how long they persist provides critical information on the dynamics of SATD management. Furthermore, by investigating the longevity and timing of SATD instances, we can better understand their impact on software quality, maintenance efforts, and the overall lifecycle of DL-enabled systems.

Based on this gap, this paper presents a preliminary mining repository study to investigate the life cycle of DL-specific SATD in order to understand (1) when and for how long developers acknowledge the DL-specific SATD and (2) what tasks developers perform when they acknowledge the presence of DL-specific SATD in DL-enabled systems. To achieve this goal, we build on top of the work done by Pepe et al. \cite{pepe2024taxonomy} by using 41 distinct types of DL-specific SATD and 40 Python open-source projects relying on TensorFlow or PyTorch. The study reveals that DL-specific SATD is predominantly introduced during the early and middle development stages, with minimal instances acknowledged in the later stages. Additionally, the lifespan of SATD varies significantly across different components. SATD related to \textit{Hardware} and \textit{Training} phases tend to last longer, while those associated with \textit{API} and \textit{Pipeline} phases are typically resolved more quickly. Specifically, most SATD were identified during feature implementation (64 instances) and bug-fixing tasks (55 instances), underscoring the prevalence of technical debt in these development activities. 
These insights highlight the nuanced dynamics of DL-specific SATD, emphasizing the need for tailored strategies to manage and mitigate effectively. 
This preliminary work is important for software engineering researchers and developers who want to understand the life cycle of DL-specific SATD and the activities performed when the SATD is recognized in DL-enabled systems, highlighting the need for tools and models to capture, track, and address DL-specific SATD across development phases and components, enabling targeted interventions. This study highlights the importance of SATD management during feature implementation and fixes, where these debts are most common. Additionally, variations in SATD duration across components underline the need for customized solutions, such as automated tools and long-term strategies to resolve technical debts. %Ultimately, this work argues for better management of technical debts in ML-enabled systems to improve maintainability and reliability.

\section{Related Work} \label{sec:RW}
The management of SATD in DL-enabled systems is increasingly crucial due to the inherent complexities and distinctive challenges these systems present. OBrien et al. \cite{obrien202223} investigate SATD in ML-enabled systems by analyzing 68,820 SATD comments from 2,641 Python-based ML repositories. Their study proposes a taxonomy of 23 ML-specific SATD types, focusing on qualitative aspects such as awareness, modularity, readability, and performance, mainly covering shallow ML projects. 
Liu et al. \cite{liu2020using} analyze SATD in seven popular DL frameworks, identifying 7,159 SATD comments through manual classification and categorizing them into seven types: design, defect, documentation, requirement, test, compatibility, and algorithm debts, with design debt being the most prevalent. Building on this work, Liu et al. \cite{liu2021exploratory} further explore the introduction and removal patterns of these debt types, finding that design debt is not only the most common but also the fastest to be resolved, offering deeper insights into its lifecycle.
Bhatia et al.\cite{bhatia2023empirical} conducted a study investigating SATD in 318 Python ML projects compared to 318 non-ML Python projects. Their findings revealed that ML projects exhibit 2.1 times more SATD, primarily due to frequent code replacements and experimental development. This research expands the existing SATD taxonomies to incorporate specific types related to ML, maps SATD occurrences to various stages of the ML pipeline, and employs survival analysis to examine the dynamics of SATD over time. While previous research focuses on shallow ML, Pepe et al. \cite{pepe2024taxonomy} created a taxonomy of 41 distinct types of DL-specific SATD analyzing 40 DL-projects based on two popular frameworks, Pytorch and TensorFlow.

Acknowledging the work done in the literature to improve knowledge about ML/DL-specific SATD, our work advances the technical debt field by analyzing when and how developers recognize SATD and the tasks performed when developers acknowledge SATD presence. This provides insight into its evolution through different stages of development and components. Unlike previous studies that mainly classify SATD types in traditional or ML-enabled systems, our research focuses on DL-specific SATD. This work contributes new perspectives on the interaction between SATD, development practices, and system quality in advanced DL pipelines.
\section{Research Questions}\label{sec:research_questions}

% Research question (QGM)
% Dataset Selection (dataset, motivazioni + descrizione)
% SATD Selection (selezione dei SATD e le motivazioni)
% Commit Extraction/Execution
% Data Analysis
To investigate when and how developers acknowledge the presence of SATD, our study adopts the Goal-Question-Metric (GQM) approach \cite{caldiera1994goal}. 
\goal{\textbf{Analyze} the lifecycle of DL-specific SATD \textbf{for the purpose of} investigating when and how SATD  is acknowledged \textbf{from the point of view} of developer and researchers \textbf{in the context of} DL-enabled system development and maintenance.}

%The goal of this study is to \textit{analyze} the presence and lifecycle of DL-specific SATD with the \textit{purpose} of understanding when and how SATD are acknowledged and the associated developer tasks in the affected files. The \textit{perspective} is of researchers and developers in the \textit{context} of ML-enabled systems.

In particular, researchers are interested in the life cycle of DL-specific SATD to develop approaches for identifying, analyzing, and mitigating DL-specific SATD. Developers are interested in the analysis of the SATD to understand how to reduce risk and maintain system quality.
%Our study aimed to identify characteristics and properties that shed light on developers' approaches when dealing with Self-Admitted Technical Debts (SATD). Specifically, we sought to understand how developers interact with SATD over time, focusing on key aspects such as when developers first recognize these issues, the typical lifespan of SATD, and whether the presence of SATD influences the change-proneness of the affected files. Our goal was to extract insights into the patterns of SATD recognition, their temporal dynamics, and the extent to which they correlate with file modifications.
To achieve our research goal, we focus on two main research questions \textbf{(RQs)} that guide our study. The \textbf{(RQ$_1$)}, seeks to identify when developers most frequently signal SATD during the software lifecycle (e.g., hardware, data, or training) and how long these issues remain unresolved. Understanding when SATD were first recognized provides insight into developers' behavior in managing technical debt and their interest in resolving it.
\begin{table*}[t]
\caption{Definition of SATD categories.}
\centering
\footnotesize
    \begin{tabular}{|l|p{0.44\textwidth}|p{0.4\textwidth}|}
    \rowcolor{arsenic}
    \textcolor{white}{\textbf{Category}} & \textcolor{white}{\textbf{Description}}& \textcolor{white}{Example}\\
    \hline
    Hardware & This category accounts for SATD dealing with hardware components, e.g., GPUs or TPUs, used for training the DL model or during inference once the trained model is deployed in a production environment. &\texttt{TODO: do something with loras and offloading to CPU} \\
    \hline
    \rowcolor{gray10}API & This category accounts for SATD due to the usage and integration of DL frameworks. &  \texttt{TODO: support native HF versions of MusicGen.}\\
    \hline
    Data & This category groups SATD related to the format/shape of the data used to train a model. & \texttt{TODO: maybe have a cleaner way to cast the input (from `ImageProcessor` side?)}\\
    \hline
    \rowcolor{gray10}Model & This category includes SATD concerning the design and setting of the DL model. & \texttt{TODO: attention mask is not used}\\
    \hline
    Training & This category encompasses SATD related to the training process, including the selection of loss functions, initialization of model parameters, learning strategies, and suboptimal implementation of training logic. & \texttt{TODO: Retrieve the seeds from the model definition instead.}\\
    \hline
    \rowcolor{gray10}Inference & This category includes SATD related to inappropriate post-processing of the outcome generated when running the trained model, as well as using suboptimal prompting. & \texttt{TODO: add prompts to .yaml}\\
    \hline
    Pipeline & This category groups SATD dealing with the setting of the DL pipeline in terms of its design and optimization. & \texttt{TODO: enable higher-order gradients}\\
    \hline
    \end{tabular}
\label{table:taxonomy}
\end{table*}
%Understanding when SATD are first recognized and how long they persist is essential to understanding the SATD lifecycle within software projects (e.g., development, feature addition, or maintenance). 
%The goal of this research question is to analyze the timing and duration of SATD acknowledgment in ML-enabled systems. It seeks to identify when developers most frequently signal SATD during the software lifecycle (e.g., development, feature addition, or maintenance) and how long these issues remain unresolved. Understanding these patterns provides insights into developer behavior in managing technical debt, understanding if the developers interest is high in resolving SATD in ML projects.
In particular, we ask:
\rqquestion{1}{When and for how long do developers acknowledge DL-specific SATD in DL-enabled systems?}
%The \textbf{(RQ$_1$)}, aim to determine when and for how long developer acknowledge DL-specific SATD, as cataloged by Pepe et al. \cite{pepe2024taxonomy}, in the development of ML-enabled systems.  
After assessing the lifecycle of DL-specific SATD, the \textbf{(RQ$_2$)} aims to identify the specific actions that developers take when they recognize DL-specific SATD.
These insights provide an understanding of whether SATD arises during feature development, bug fixing, or other activities. 
%This research question investigates the specific actions developers take when acknowledging SATD in ML-enabled systems. It explores whether they mark SATD with comments (e.g., "TODO," "FIXME"), make partial changes, or add new functionalities in the ML-enabled system. It also examines whether SATD arise during feature development, bug fixes, or other tasks. The goal is to understand developer practices and the context of SATD acknowledgment, offering insights into how technical debt is identified and managed in ML projects.
\rqquestion{2}{What tasks do developers perform when acknowledging the presence of DL-specific SATD in DL-enabled systems?}
% Finally, the \textbf{(RQ$_3$)} focuses on how the presence of SATD affects the nature of changes to affected files and the possibility of introducing problems into the code in the long development of ML-enabled systems.
% %This question focuses on how the presence of SATD impacts the frequency and nature of changes to affected files, helping us understand their role in driving or hindering codebase evolution.
% \rqquestion{3}{What is the change-proneness of files affected by specific SATD in ML-enabled systems?}
% %Our research method include the application of a Mining Software Repositories (MSR) study to collect historical information from commits and files affected by SATD.
To design our process and analyze the results of our mining study, we followed the guidelines specified in the “Repository Mining” and “General Standard” categories of the ACM/SIGSOFT Empirical Standards \cite{ralph2020empirical}.
%Additionally, to define our design process, we follow the guidelines of Codabux et al. \cite{codabuxteaching}.  
%\todo[inline]{G: Explain the steps of the process of the analysis.}

\section{Research Method}\label{sec:research_method}
This section describes our study research approach to answer our  \textbf{(RQs)}. %Figure \ref{} provides the main steps conducted.

\subsection{Dataset Description}
%\todo[inline]{Insieme: prima descriviamo i progetti, poi gli SATD e le relative categorie con tabella.}
The first step is to select appropriate projects within DL-specific SATD for our analysis. For this purpose, we built our study upon the work of Pepe et al. \cite{pepe2024taxonomy} for a twofold reason.

On the one hand, the contribution of Pepe et al. \cite{pepe2024taxonomy} offers a curated dataset of 100 ML open-source projects, with 46  affected by SATD, providing a strong basis for our study. These projects are carefully selected to ensure relevance, focusing on Python-based repositories utilizing the two well-known DL frameworks, PyTorch\footnote{\url{https://pytorch.org/}} and TensorFlow\footnote{\url{https://www.tensorflow.org}}. This dataset was refined by excluding tutorials, toy projects, and repositories not primarily written in Python, ensuring the inclusion of active, high-quality, and representative projects. Moreover, the dataset guarantees that all projects included contain at least one DL-specific SATD. 

On the other hand, Pepe et al. \cite{pepe2024taxonomy} contribute a comprehensive taxonomy of DL-specific SATD that classifies and contextualizes the technical debts found in ML projects. This taxonomy, in Table \ref{table:taxonomy}, classifies 200 SATD instances into 41 distinct types specific to DL.
The types are further grouped under seven main categories, referring to aspects such as hardware configurations, training processes, data handling, and inference issues, all of which are essential in the lifecycle of DL-enabled systems.
%grouped into two high-level domains: Infrastructure and DL life-cycle.
%These categories include aspects such as hardware configurations, training processes, data handling, and inference issues, all of which are essential in the lifecycle of ML-enabled systems. %The taxonomy serves as a valuable reference for understanding technical debt in ML systems and offers insights into common challenges faced by developers.

%To validate the continued relevance of the SATD instances in the dataset, we performed a confidence check, \ie checking the existence of the corresponding repositories on GitHub affected by SATD instances at the date of the study.
To maintain the relevance of the SATD instances in the dataset, we conducted a verification process to confirm the data's reliability and validity. Specifically, we verified whether the GitHub repositories associated with the identified SATD instances were still accessible and actively maintained at the time of our study. %This step ensures that the dataset accurately reflects current and applicable software development practices.
Specifically, for each of the SATD instances analyzed with the validated dataset, 
we relied on Github API \cite{GitHubREST} to ensure the existence of the repository and the related commit for which the SATD has been identified. This step guarantees that all the SATD selected as the starting set of experimental objects are available to extract the commit history.
Therefore, after this step, we selected a total of 185 SATD instances across 40 GitHub projects. 
%These projects relied on two widely-used ML-specific libraries, which played a central role in determining the suitability of the repositories for inclusion in the initial dataset.
By focusing on this refined subset, we ensured that the SATD instances and their corresponding projects were relevant to our study’s objective of analyzing technical debt in DL-enabled systems.
%The first step is to select appropriate projects for our analysis within DL-specific SATD. We select the work done by Pepe et al. \cite{pepe2024taxonomy} for a twofold reason. On the one hand, the contribution of Pepe et al. \cite{pepe2024taxonomy} contains a curated dataset of 100 ML open-source projects. In particular, this dataset contains Python projects made with two well-known ML development libraries, PyTorch and Tensorflow, and contains projects with al least one DL-specific SATD. This starting dataset provided a solid basis for our study by narrowing the scope to projects that were both active and representative of ML-enabled systems. On the other hand, the authors have created a catalog of DL-specific SATD relevant to the context of ML development.

%To collect historical information on SATD, we aimed to identify a comprehensive set of these issues as a foundation for our analysis. The primary goal was to ensure the dataset captured relevant and actionable SATD instances reflective of real-world software development scenarios.
%We began with the dataset of SATD released by Pepe et al. \cite{pepe2024taxonomy}, which was specifically curated based on predefined selection criteria. These criteria ensured the dataset focused on projects focusing on two well-known ML development libraries (PyTorch and Tensorflow) and contained meaningful SATD instances relevant to the context of machine learning development. This starting dataset provided a robust foundation for our study by narrowing the scope to projects that were both active and representative of ML-enabled systems.

\subsection{History Mining}

\begin{figure}
    \centering
    \includegraphics[width=\linewidth]{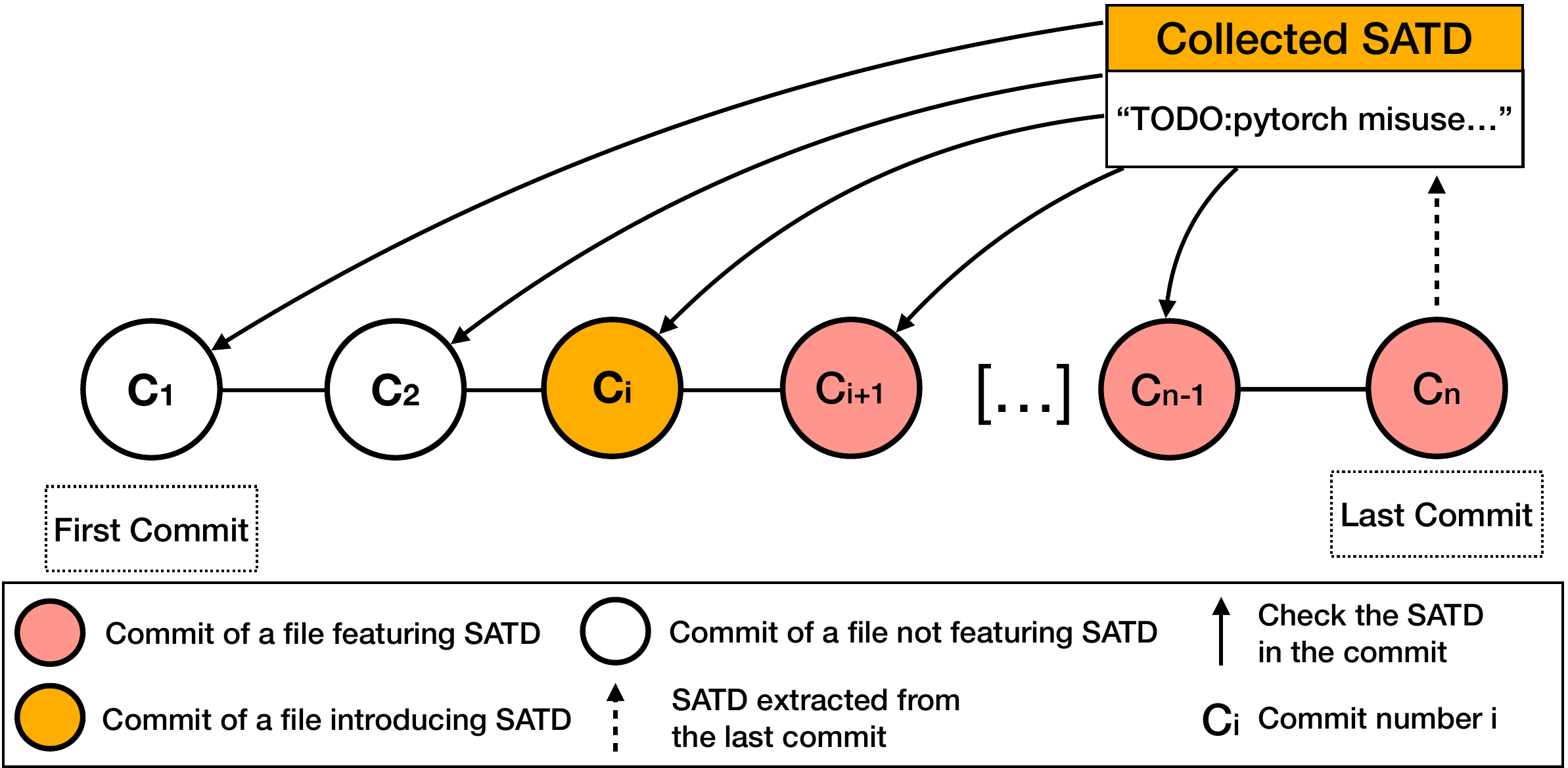}
    \caption{SATD introducing commit extraction}
    \label{fig:intro-commit}
\end{figure}
After selecting the project dataset and the catalog of DL-specific SATD, we analyzed the commits in Github projects. In particular, we proceeded to the commit extraction phase to identify the specific commits where each DL-specific SATD was first introduced into the codebase. To achieve this, we utilized PyDriller\footnote{\url{https://github.com/ishepard/pydriller}}, a Python framework designed for mining Git repositories and analyzing commit histories. PyDriller allowed us to programmatically retrieve detailed information for each commit, including the commit hash, the lines modified, and the associated commit message. This simplified the data extraction process and ensured a systematic approach to managing the large volume of commits in projects.
To determine the specific commit that introduced each DL-specific SATD instance, we employed a reverse tracing strategy similar to that performed by Tufano et al. \cite{tufano2015and}, commonly used for analyzing the history of code smells. As depicted in Figure \ref{fig:intro-commit}, 
the process began by leveraging the last commit to extract the SATD instance and to trace backward through the commit history.
%the process began by identifying the last active commit where the SATD instance was confirmed to exist and traced backward through the commit history.
Leveraging the DL-specific SATD annotations, we systematically traversed the commit history in reverse chronological order, examining earlier commits to locate the exact introduction point of the DL-specific SATD instance. This backward traversal method ensured accurate mapping between DL-specific SATD instances and their origin commits.

When analyzing the commit history of a project $P$. Let a file $F$ in a project $P$ identify the introduction of a specific $s$ $\in$ DL-specific SATD. Two scenarios can arise when encountering a commit that modifies the $F$ file and the prior version of $F$ does not include the specified DL-specific SATD:
\begin{enumerate}
    \item \textbf{Last occurrence but not the first:} This commit marks the most recent occurrence of the $s$ $\in$ DL-specific SATD in the file. However, it does not necessarily represent the first introduction of the DL-specific SATD, as it might have been added earlier in the commit history.
    \item \textbf{First occurrence:} If no earlier versions of the $F$ file, as examined through the commit history, contain the $s$ $\in$ DL-specific SATD, then this commit is identified as the one that introduces $s$. In this case, it is considered the introducing commit for the DL-specific SATD.
\end{enumerate}
%In addition, there is the possibility of reintroduction: a $s$ $\in$ DL-specific SATD that was previously resolved may reappear in a later commit due to subsequent changes. To account for this, we kept going back through the commit history to identify multiple instances of DL-specific SATD introduction.

This approach ensures that all occurrences of $s$ $\in$ DL-specific SATD, whether it is the first occurrence or not, are accurately recorded and analyzed.
%As soon as the current commit under analysis modified the file $F$ and the previous version of $F$ does not contain the specified SATD, two conditions can occur. Firstly, this is for sure the last occurrence for which the SATD $s$ is encountered, but not necessarily the first one. If none of the versions of the file $F$ analyzing the commit history does not contain the specified SATD s,
%this is also the first commit in which occur the SATD, resulting in the introducing commit of the SATD s.
%Otherwise, there could be the possibility for which an SATD has been reintroduced after a modification, recognizing a SATD s already recognized in a previous commit. Therefore, we continued through the commit history to extract multiple occurrences of the introduction of SATD.  
%After collecting all the versions of the files through the commit history that contain the specified SATD, we defined the introducing commit $c_i$ for the SATD $s$ if the SATD annotation was present in $c_i$ but absent in all preceding commits.
Specifically, we defined the introduction commit $c_i$ for the $s$ $\in$ DL-specific SATD as the first commit in which the DL-specific SATD annotation appeared. Using this approach, we identified commits that introduce SATD across all file versions in the commit history, collecting a total of 5,337 commits.
%Therefore, applying this approach, we identified the SATD-introduction commits among all the file versions in the commit history, analyzing a total of 5337 commits.

\subsection{Data Analysis}
The following section describes our approach to analyzing the data collected to answer our research questions.
%Starting from the SATD-introducing commit, we analyzed properties and characteristics of SATD to address each of our research questions, as described in the following.
%\textit{RQ1: When and why do developers signal the presence of specific SATD in ML-enabled systems?}
To address \textbf{(RQ$_1$)}, we conducted a comprehensive analysis focusing on temporal data of the commit history associated with DL-specific SATD. Specifically, we developed a Python script that writes this information into a \texttt{csv} file. Once a $s$ $\in$ DL-specific SATD is found in an $F$ File, $s$ is saved, and the commit history of $F$ is traversed backward until the introduction of $s$. 
Additionally, in the saved dataset, we extract the current commit date and the previous commits, collecting the whole history for each file featuring the presence of a DL-specific SATD instance. Moreover, through the backward tracing of the commit history, we record the line number for every commit affected by a specific SATD instance to control the presence of the SATD also when the file changes along the history.

\textbf{When SATD is introduced.} To determine the introduction of any $s$, we measured the time and number of commits from the start of the project to the most recent commit, where it was still present.  %To calculate the lifetime of any $s$, the time and number of commits between their introduction and the most recent commit in which they were still present were measured, highlighting how long developers leave $s$ unresolved.
To analyze the introduction period of DL-specific SATD in their project lifetime, we examined their position in the commit history. Specifically, we calculated the difference in the number of commits between the repository's first commit and the commit that introduced DL-specific SATD. This analysis allows us to identify at what point in the project developers had reported $s$, measured by the number of commits made prior to its introduction.
%To analyze the position of SATD-introduction commits in the development lifecycle of a project, we examined their placement within the commit history relative to the total number of previous commits. This analysis aimed to determine how many commits had been made before SATD was reported by developers from the start of the project. 
Collecting all SATD introduction positions over the project lifetime, we segmented the commit history of each project into three phases:

\begin{itemize}
	\item \textbf{Early Stage}: The first 25\% of commits in the project timeline.
	\item \textbf{Middle Stage}: Commits between 25\% and 75\% of the project timeline.
	\item \textbf{Last Stage}: The final 25\% of commits in the project timeline.
\end{itemize}
This segmentation provided a perspective that highlights overarching insights on when DL-specific SATD-introduction commits occur within a project's lifespan.

\textbf{How long SATD persists.} Additionally, to examine the persistence of DL-specific SATD instances, we evaluated how long they remained unresolved by calculating the number of days between their introduction and the latest commit in the repository. It is important to note that the DL-specific SATD dataset used in this study includes only DL-specific SATD instances that are still present in the project's codebase. These instances were identified from the last active commit and traced backward through the commit history.

This analysis made it possible to determine how long developers continued to work on the project while leaving DL-specific SATD instances unresolved. Indicating the length of time DL-specific SATD instances were present in the system until the latest project version.

Combining the previous two analyses allows us to obtain information on commits that introduce SATD over the project lifetime, highlighting when they are introduced and how long they persist, answering to \textbf{RQ$_1$}.

%In particular, starting with the introducing commits,  we developed a Python script that calculated the time since the first project commits to the $s$ SATD introduction to identify whether SATD were typically added during early development phases, active feature additions, or later maintenance stages. These were categorized into three phases: initial development (first 25\% of the timeline), active development (25\%-75\%), and maintenance (last 25\%). Additionally, the lifespan of SATD was analyzed by measuring the time and number of commits between their introduction and the most recent commit where they were still present, highlighting how long developers leave SATD unresolved.

%\textit{RQ2: What tasks do developers perform when acknowledging the presence of specific SATD in ML-enabled systems?}

To address \textbf{(RQ$_2$)}, %to analyze the tasks developers perform when acknowledging SATD in ML-enabled systems, 
we focused on examining the commit messages associated with the introducing commits. Commit messages were analyzed using keyword-matching strategies to identify activities in which developers acknowledged DL-specific SATD. Based on the presence of keywords, the analysis categorized code modifications into four main types: feature development, bug fixes, enhancement, and refactoring. 

The selection of these types has been found to be effective in classifying commit messages \cite{tufano2015and}. 
By using keyword-matching strategies, we systematically categorize SATD acknowledgment specific to DL in commit messages and analyze the activities that lead developers to indicate technical debt.
%However, this strategy has not been successful in identifying all the categories for all the commits.
Although this strategy helped us identify some categories, it did not allow for the identification of all categories.
In detail, starting from a set of 185 DL-specific SATD commits, we extracted 62 commits under the four starting categories through keyword matching. 

Therefore, the first two authors manually analyzed the remaining set of 123 commits by conducting content analysis
 \cite{krippendorff2018content}. Every inspector reviewed every comment, trying to classify it into one of the categories predefined previously. In case of disagreement, the two inspectors analyzed the results and started a discussion to reach an agreement.
After this phase, 108 commits were manually categorized into the categories of Table \ref{table:taxonomy}, still leaving 15 cases of commits introducing DL-specific SATD uncategorized. 
Finally, a new round of inspection on the remaining part has been done to identify new categories among the meaningful commits. Therefore, a new category, named "chore", has been identified in addition.

Specifically, 11 out of the 15 commits are grouped into this category, identifying commit operations that aim to perform maintenance tasks that do not directly affect the software's functionality or features, such as updating dependencies or improving documentation.

Therefore, we analyzed the following categories: 

\begin{enumerate}
    \item \textbf{Bug fixes:} where DL-specific SATD were acknowledged during defect resolution or maintenance tasks, identified by keywords such as \textit{fix} or \textit{bug};
    \item \textbf{Enhancement:} where DL-specific SATD were acknowledged during the improvement of existing features or functionalities, often identified by keywords such as \textit{improve} or \textit{enhance};
    \item \textbf{Feature development:} where DL-specific SATD were introduced alongside the implementation of new functionalities, identified by keywords such as \textit{feature} or \textit{add}; 
    \item \textbf{Refactoring tasks:} where DL-specific SATD were tied to incomplete or deferred structural changes, often marked by keywords like \textit{refactor}. 
    \item \textbf{Chore:} where DL-specific SATD were tied to routine maintenance, such as dependency updates or configuration changes, identified manually. 
\end{enumerate}
Four commits introducing DL-specific SATD were not categorized and had no meaningful description; to avoid possible misclassification, we excluded these commits from our analysis. 
Finally, we classified 181 SATD-introducing commits into their specific five categories.
Collecting the occurrences of commit categories that introduce allows us to investigate what are the activities that lead developers to recognize and denote DL-specific SATD, answering to \textbf{RQ$_2$}.
The entire process to reproduce our work is in the online appendix \cite{appendix}.

%Finally, to address \textbf{(RQ$_3$)}, we analyzed the change-proneness of files affected by specific SATD in ML-enabled systems, we examine the commit history of these files to identify bug fix commits both before and after the SATD introduction. Using introducing commits as reference points, we categorize bug fix commits based on their timing relative to the SATD introduction. Bug fixes are identified through keywords such as fix: or bug: in commit messages. By comparing the frequency and nature of bug fixes across these periods, we assess whether SATD presence correlates with increased bug-proneness.  This approach provides insights into the relationship between SATD and file stability.

\section{Results}
\label{sec:results}
In this section, we report quantitative insights from the repository mining study to address the \textbf{RQs}:

\subsection{RQ$_1$: When and for how long do developers acknowledge DL-specific SATD in DL-enabled systems?}
\begin{figure}
    \centering
    \includegraphics[width=\linewidth]{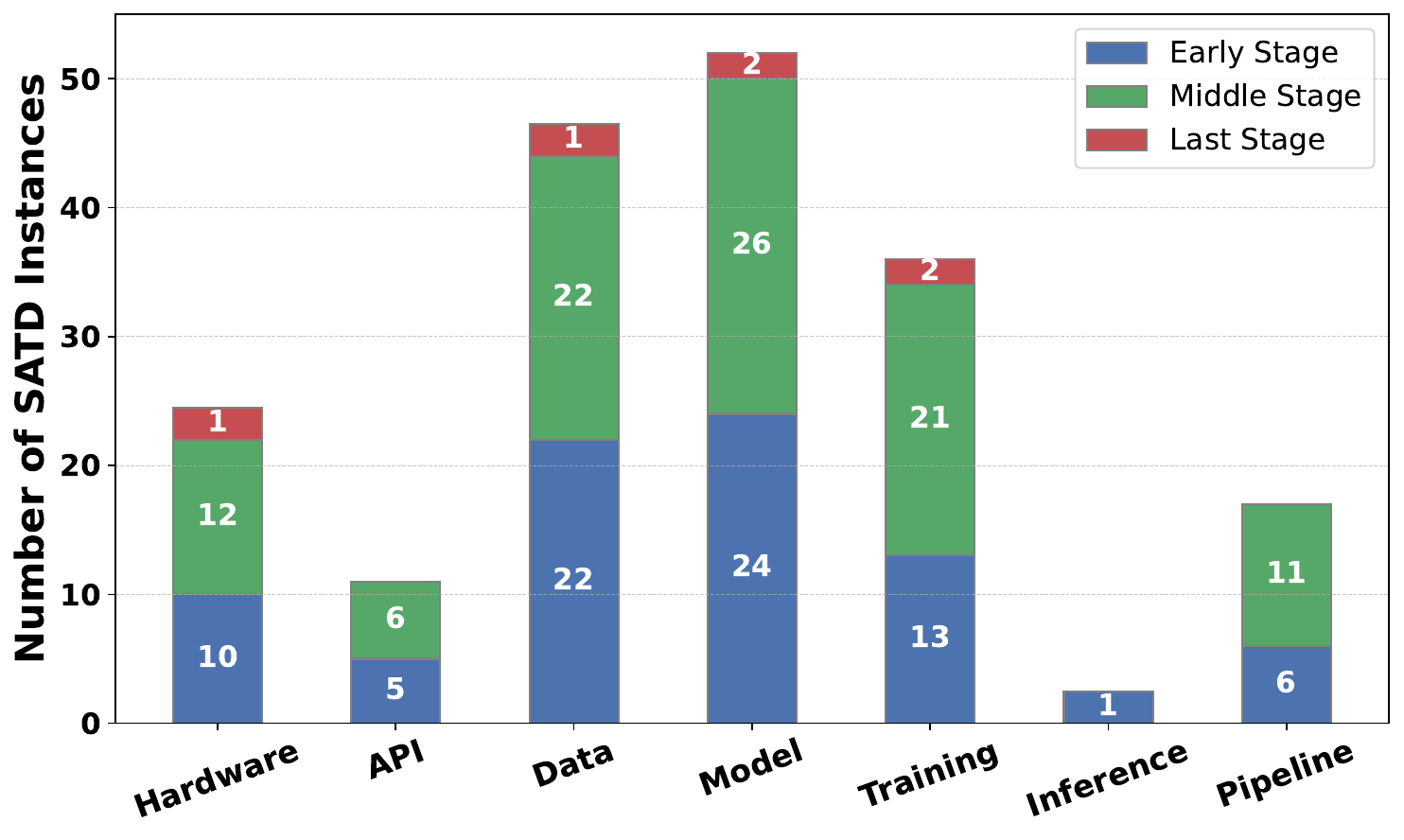}
    \caption{Frequency of SATD for each development stage.}
    \label{fig:rq1_frequency}
\end{figure}
Figure \ref{fig:rq1_frequency} presents the boxplot of the distribution of DL-specific SATD instances across the early, middle, and last stages of development for various components within DL-enabled systems across 40 projects. The analysis reveals that DL-specific SATD is most frequently acknowledged during the middle stage (98), followed by the early stage (81), and only minimally in the last stage (6). This trend underscores that developers primarily recognize and address technical debt early and during the refinement phases of the lifecycle.

Considering the totality of all projects, the \textit{Model} phase has the highest total D-specific SATD instances (52), with 26 cases in the middle stage and 24 instances in the early. Only two are in the last stage. The \textit{Data} phase is the second most frequent (45), with 24 in the middle stage, 22 instances in the early stage, and only one in the last stage.

The \textit{Training} phase follows, with a total of 36 instances, distributed in the middle stage with 21 instances, while the early stage accounts for 13 instances, and only 2 are acknowledged in the last stage. The \textit{Pipeline} phase shows 17 instances in total, with 11 occurring in the early stage and 6 in the middle stage, with no debt reported in the last stage. In contrast, the \textit{Hardware} phase accounts for 22 instances, mainly concentrated in the middle stage (12 instances) and the early stage (10 instances), with none reported in the last stage. The \textit{API} phase has the lowest number of instances among the major phases, with a total of 11 cases: 5 in the early stage, 6 in the middle stage, and none in the last stage. Finally, the \textit{Inference} phase shows only one instance in the early stage. 

The findings reveal a trend where most DL-specific SATD is reported in the early and middle stages, with only residual debt persisting into the last stage across most components.

\begin{figure}
    \centering
    \includegraphics[width=0.9\linewidth]{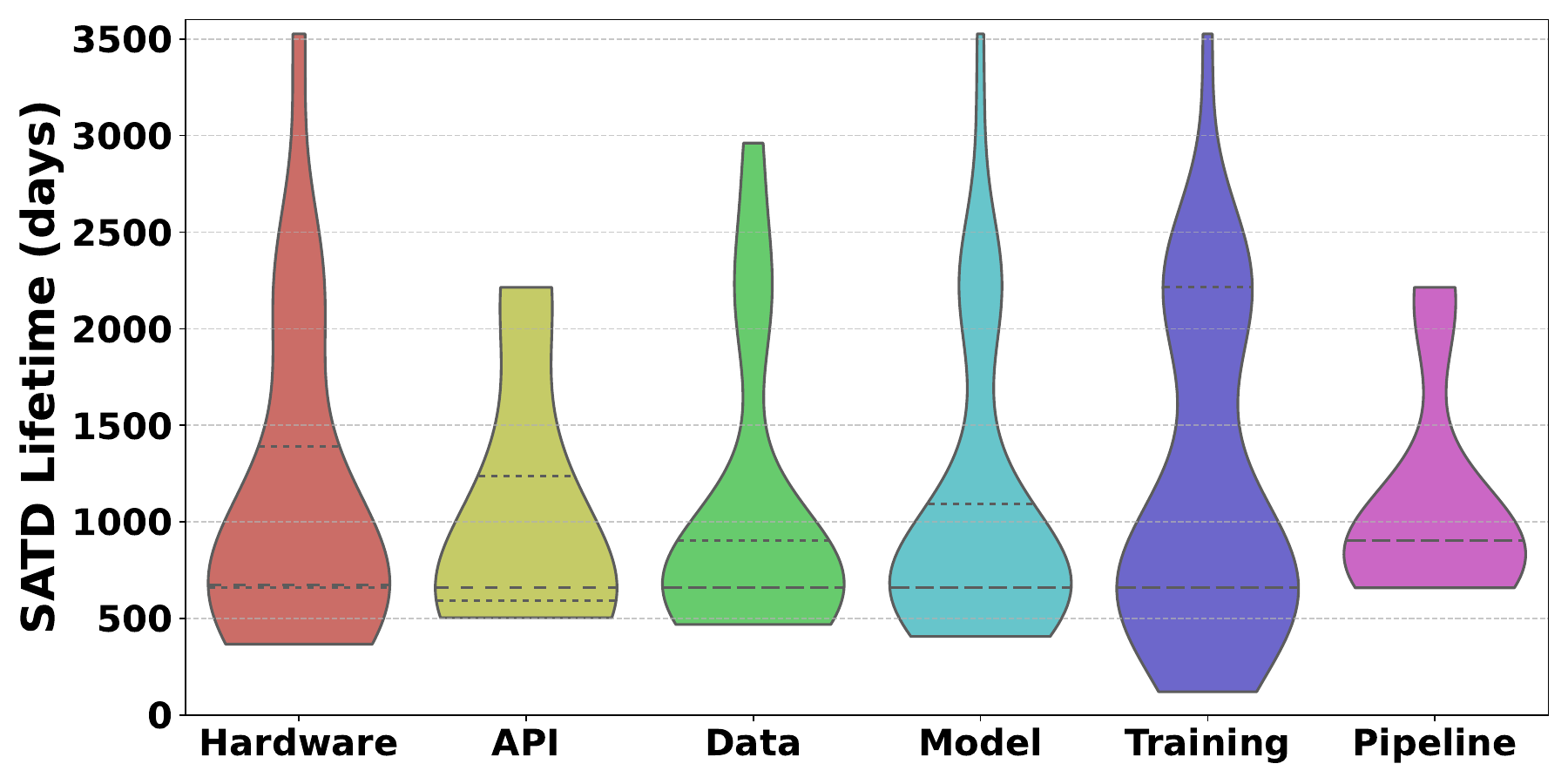}
    \caption{Distribution of SATD Lifetime since their introduction.}
    \label{fig:rq1_how_long_violin}
\end{figure}

The violin plot shows in Figure \ref{fig:rq1_how_long_violin} the distribution of SATD lifetimes, measured in days, across various components of the DL-enabled system. The diagram reveals the variability in the persistence of DL-specific SATD across components. For example, \textit{Hardware} and \textit{Training} phases exhibit the broadest distributions, with some instances persisting for extended periods, indicating that technical debt in these areas can remain unresolved for years. The median lifetimes for these components suggest that resolving such debt is often delayed, likely due to the complexity or lower prioritization of these issues. In contrast, components such as \textit{Pipeline} and \textit{API} phases exhibit a constrained distribution with shorter lifetimes. This indicates that DL-specific SATD in these areas is typically addressed more promptly and consistently. Meanwhile, \textit{Model} and \textit{Data} phases fall somewhere in between, with DL-specific SATD lifetimes showing moderate distributions and occasional instances of long-lasting debt. This reflects a mix of challenges that are resolved at varying rates. Overall, the plot highlights the differing persistence of DL-specific SATD across components, emphasizing the need for targeted strategies to tackle long-lived technical debt.

\rqanswer{1}{Developers mainly identify DL-specific SATD during the early (81 instances) and middle (98 instances) stages, with few instances (6) in the final stage. SATD is most common in the \textit{Model} and \textit{Data} and less in the \textit{API} and \textit{Inference} stages. 
The duration of SATD varies significantly, with the \textit{Hardware} and \textit{Training} phases having the longest-lasting issues. In contrast, the \textit{Pipeline} and \textit{API} phases are resolved more quickly. Generally, developers address most SATD by the refinement phase, although some problems persist longer in certain components.}

\subsection{RQ$_2$: What tasks do developers perform when acknowledging the presence of DL-specific SATD in DL-enabled systems?}
\begin{table}[h!]
\centering
\renewcommand{\arraystretch}{1.2} % Adjust row height
\setlength{\tabcolsep}{8pt} % Adjust column spacing
\arrayrulecolor{black} % Set rule color to black

\begin{tabular}{|l|p{0.05\columnwidth}|c|c|c|c|c|}
\hline
\rowcolor{arsenic}
\textcolor{white}{\textbf{Category}} & \textcolor{white}{\textbf{B}} & \textcolor{white}{\textbf{F}} & \textcolor{white}{\textbf{E}} & \textcolor{white}{\textbf{C}} & \textcolor{white}{\textbf{R}} & \textcolor{white}{\textbf{Total}}\\ 
Hardware   & 7  & 5  & \textbf{8}  & 1 & 1 & 22 \\ 
\hline
API        & 3  & \textbf{4}  & 2  & 0 & 1 & 10\\ 
\hline
Data       & \textbf{16} & 15 & 8  & 5 & 0 & 44\\ 
\hline
Model      & 14 & \textbf{20} & 14 & 1 & 2 & 51\\ 
\hline
Training   & 6  & \textbf{15} & 6  & 2 & 7 & 36\\ 
\hline
Inference  & 0  & \textbf{1}  & 0  & 0 & 0 & 1\\ 
\hline
Pipeline   & \textbf{5}  & 4  & 3  & 2 & 3 & 17\\ 
\hline
\rowcolor{gray10}\textbf{Total} & 55 & \textbf{64} & 41 & 9 & 12 & 181\\
\hline

\end{tabular}

\caption{Distribution of SATD Types Across Categories. Bug Fix (B), Feat (F), Enhance (E), Chore (C), Refactor (R)}
\label{tab:satd_distribution}
\end{table}

%overview, senza le categorie, chi viene introdotto prima in total?
Table \ref{tab:satd_distribution} reports the occurrences of the commit messages that introduce the DL-specific SATD, grouped in the seven categories. Analyzing the commit messages, we found a high occurrence (64 instances) of DL-specific SATD introduced during the \textit{implementation of a new feature} operation. Still, many DL-specific SATD are consequently introduced during a \textit{bug-fixing} operation (55 instances) and operations related to the \textit{enhancement} of the system (41 instances). Conversely, a lower amount of instances are introduced during a \textit{refactoring} (12 instances) or a \textit{chore} (9 instances). 

DL-specific SATD is almost equally present across different operations categories. However, the analysis varies significantly when considering the specific types of DL-related SATD.
From the set of commit messages analyzed, we found that 8 of the 22 \textit{Hardware} related DL-specific SATD are introduced during an enhancement operation, resulting in the most recurrent in this category. 
Data-related DL-specific SATD usually appears when a \textit{bug-fixing operations appear}. For instance, in the project \textit{huggingface/diffusers}, during a bug-fixing operation related to the feature extraction component, a data-related SATD has been denoted by the developer's raising the absence of a scaling operation \cite{hf_diffusers_data_example}. 

The most recurrent operation that introduces Model-related SATD is the implementation of the new feature with 20 instances. 
From the same commit operation related to the release of a new version of structured prompting, retrieved in the project \textit{microsoft/LMOps}, we found the introduction of the developers of four Model-related SATD, raising problems related to the encoding components of transformers-based models (\eg \quoted{\textit{\# TODO: we should reuse the pretrained model dict which already has mask}}) \cite{ms_lmops_example}.

Training-related SATD is mostly introduced during the implementation of a new feature, and 15 instances were found for this operation. Also, in this case, the addition of a new feature is denoted by developers to cause problems in the components that are involved in the training phase. Specifically, a developer denoted the introduction of a dummy encoder for the Roberta model to fix the non-determinism model (\ie \quoted{\textit{\# TODO: remove after fixing the non-deterministic text encoder}}) when introducing a new component in the system \cite{hf_diffusers_training_example}.
Among the 17 instances of pipeline-SATD, we found that every commit-operation type introduces at least one SATD of this category, with a slightly higher tendency when performing bug-fixing operations, with 5 instances.
Finally, the unique SATD related to the inference of DL models was found to be introduced during the implementation of a new feature. 

In summary, there is a higher tendency to highlight  DL-specific SATD when a new feature is implemented. Moreover, four of the categories are found to be introduced during this type of commit operation. At the same time, \textit{Pipeline}, \textit{Hardware}, and \textit{API} SATD instances are occurrences that are distributed among different categories of commit operations.

%\rqanswer{2}{Developers denotes the presence of DL-specific SATD during the implementation of a new feature, identifying several categories of SATD during this operation. Moreover, some of the SATD instances, such as \textit{Data} and \textit{Hardware}, highlight their occurrence when bug fixing or enhancement operations are performed in the system.}

\rqanswer{2}{Developers more frequently introduce DL-specific SATD during feature implementation (64) and bug fixes (55). In contrast, SATD is less commonly associated with refactoring (12) or chore categories (9). DL-specific SATD introduced also varies based on the activity: \textit{Data} SATD is often encountered during bug fixes (16), while \textit{Model} and \textit{Training} SATD are typically introduced during the development of new features (20 and 15 respectively).}

\section{Threats to Validity}\label{sec:threats}
This section outlines the potential limitations and our strategies for addressing them.

\textbf{External Validity.}  One of the main challenges is the representativeness of the data sample used in our investigation of DL-specific SATD. To mitigate the threat to generalizability of the dataset, we relied on the study of Pepe et al. \cite{pepe2024taxonomy} consisting of 40 open-source Python projects using TensorFlow and PyTorch and 185 DL-specific SATD types and applied a verification process to confirm the validity of the data. Although this sample provides a basis for our investigation, it may not reflect the diversity of DL-specific SATD types across software systems and contexts, since it does not include resolved SATD instances. Therefore, replications are needed to evaluate our preliminary observations, expanding the study to a larger scale. Additional investigations allow for the identification of additional DL-specific SATD types and the discovery of patterns among them. 
An additional threat concerns the diversity of the ML Projects affected by DL-specific SATD. To mitigate this threat, the Pepe et al. \cite{pepe2024taxonomy} dataset contains only projects that have at least one SATD instance, ensuring the reliability of the dataset. Additionally, our study focuses on PyTorch and Tensorflow projects, given its widespread use.

%Factors such as the AI-technology domain (e.g., AI, web development, embedded systems) and the size and maturity of the repository are likely to influence the nature and frequency of SATD. However, broader studies can also introduce risks, such as an increased likelihood of false positives. To address this concern, our analysis specifically focused on a previously validated SATD dataset, which ensures the reliability of the experimental subjects. Nonetheless, future research could improve external validity by incorporating a more diverse dataset that includes a wider variety of projects and domains.

%\textbf{Internal Validity.} A potential threat is the incomplete extraction of SATD instances from commit histories. While we used the validated dataset by Pepe et al.~\cite{pepe2024taxonomy}, there is a possibility that some DL-specific SATD instances were missed during repository mining. To mitigate this, we employed PyDriller for systematic commit analysis and conducted additional manual verification of sampled results. Furthermore, the variability in how developers document SATD across projects could influence detection rates.

\textbf{Construct Validity.} One potential threat arises from the classification of DL-specific SATD instances based on keyword matching and manual inspection. Misinterpretation of SATD categories or commit messages could lead to incorrect classifications. To address this, we leverage the taxonomy of DL-specific SATD \cite{pepe2024taxonomy} to ensure the use of consistent definitions, and we conducted content analysis
sessions \cite{krippendorff2018content} to minimize bias in the manual categorization. Disagreements between inspectors were resolved through discussions to reach consensus. The data analysis was conducted by the first two authors, who have substantial expertise in AI software engineering. The first author is a Ph.D. student with over four years of experience in tech debt and software engineering for artificial intelligence (SE4AI). The second author is a Ph.D. student with over three years of experience in DL and SE4AI.

%Another concern involves the accuracy of commit categorization. While keyword matching offers an efficient way to identify explicit cases, it struggles with more complex or implicit instances. To address this, we combined keyword matching with manual inspection to handle uncategorized commits. Although effective, manual inspection introduces the risk of subjective bias, as categorization can vary between inspectors. To mitigate this, we implemented a dual-inspector process, where two independent inspectors categorized the commits, with disagreements resolved through discussion by the first authors. This approach balances the efficiency of automation with the reliability of manual verification, reducing subjectivity.

\section{Discussion}\label{sec:discussion}

Analysis of the prolonged persistence of DL-specific SATD revealed several discussion points.

\textbf{The need for the prevention of SATD}. Our results indicate that developers frequently recognize DL-specific SATD during the initial and middle stages of DL projects across all categories analyzed. Notably, 64 DL-specific SATD instances are introduced during feature implementation, suggesting that these issues often emerge during active development when developers prioritize meeting deadlines or delivering functionality over addressing technical debt. The long persistence of these issues throughout the project lifecycle highlights a tendency to leave acknowledged SATD unresolved. 
This behavior can lead to compounding challenges, such as reduced system maintainability and potential performance degradation. Addressing SATD early, especially during feature development, is crucial to mitigating these risks and ensuring the long-term sustainability of DL-enabled systems \cite{mcguire2023sustainability,sculley2015hidden}. 
 
While DL-specific SATD exhibits unique characteristics, such as challenges in data dependencies, model interpretability, and retraining, our findings suggest similarities with traditional SATD in its upward trend and persistence \cite{bavota2016large}. This implies that established techniques like code reviews or static analysis could be adapted, although further research is needed to tailor them for DL-specific challenges. Persistent DL-specific SATD can also have broader implications, including risks to system reliability, especially in critical domains.

\textbf{Combining the complexity of DL with SATD.}
Although this analysis focuses on the timing of SATD introduction and does not delve into resolution strategies, the long persistence found in these projects has some implications. 
Analyzing SATD in traditional software systems, Potdar and Shihab \cite{potdar2014exploratory}  discovered that even after many releases, only less than 64\% percent of SATD instances are resolved. On the same line, Bavota and Russo \cite{bavota2016large} also analyzed the lifespan of traditional SATD instances, discovering that these self-acknowledged issues remained unresolved for more than 1000 commits since the start of the project. 
%The analysis conducted by Bhatia et al. \cite{bhatia2023empirical} revealed that SATD in ML projects emerges earlier and persists longer than in traditional systems, especially in critical stages like data preprocessing and model training.
In the context of DL technologies, we found that many of the DL-specific SATD instances considered in this study, especially for the \textit{training-related} category, remained unresolved for longer periods. 
Thus, the complexity of managing multiple components, such as pipelines, model architectures, and training processes, combined with the difficulty of dealing with SATD in traditional software systems, suggests that developers face significant hurdles in solving problems that may propagate not only on the current state of the system but also on future iterations within specific DL-systems. This highlights the need for new practices to address the unique challenges faced in DL projects. Without such strategies, the accumulation of unresolved SATDs could intensify the inherent complexity of these systems, threatening their maintainability over time. Therefore, to support developers in managing SATD, there is a pressing need to develop tools tailored to DL-specific contexts. These could include automated SATD tracking integrated into ML pipelines, enhanced visualization of debt accumulation within lifecycle stages, and prioritization mechanisms that quantify technical and operational risks associated with unresolved SATD.

%The long persistence of SATD in ML systems also reveals a gap in current tooling and practices. Existing SATD detection and management frameworks are largely optimized for traditional software systems and do not account for the unique characteristics of ML workflows. To address this, there is a pressing need to develop tools and methodologies tailored to DL-specific contexts. These might include automated SATD tracking integrated into ML pipelines, enhanced visualization of debt accumulation within model lifecycle stages, and prioritization mechanisms that quantify the technical and operational risks associated with unresolved SATD.

\section{Conclusion and Future Work} \label{sec:conclusions}
In this study, we analyzed the occurrence and evolution of DL-specific SATD over time in DL-enabled systems to better understand when and how developers acknowledge these quality issues.
By analyzing 40 open-source projects and 185 DL-specific instances, our investigation reveals critical insights into the lifecycle of DL-specific SATD. Notably, we observed that DL-specific SATD is mainly acknowledged during the early and middle stages of the DL software lifecycle, often persisting throughout the project’s development. Our analysis also revealed a high prevalence of DL-specific SATD instances introduced during the implementation of new features, indicating that DL-specific SATD often emerges when new production code is added to the system.

These two outcomes provide a foundation for investigating DL-specific SATD to determine if and how its persistence after adding new system components leads to quality issues, harming the maintenance process.
Therefore, future work will focus on further investigations to evaluate the practical motivations of DL-specific SATD persistence from the developer's perception. 
Moreover, we aim to expand the dataset to include a larger set of instances, enabling their identification that was acknowledged in the past and subsequently resolved. This extended analysis will facilitate the discovery of common patterns and trends, paving the way for the development of refactoring techniques for DL-specific SATD.

\section*{Acknowledgments}
This work has been partially supported by the European Union - NextGenerationEU through the Italian Ministry of University and Research, Projects PRIN 2022 "QualAI: Continuous Quality Improvement of AI-based Systems" (grant n. 2022B3BP5S, CUP: H53D23003510006).
\balance
\bibliographystyle{IEEEtran}
\bibliography{Paper.bib}

\end{document}